\documentclass[sigconf,nonacm]{acmart}

\setcopyright{none}
\renewcommand\footnotetextcopyrightpermission[1]{}
\acmDOI{}
\acmISBN{}

\usepackage{booktabs}
\usepackage{balance}
\usepackage{xspace}

\newcommand{\vh}{\textsc{VeriHarness}\xspace}
\newcommand{\rawdiag}{\textsc{RawDiag}\xspace}
\newcommand{\samenl}{\textsc{SameNL}\xspace}
\newcommand{\locobs}{\textsc{LocObs}\xspace}
\newcommand{\typedfields}{\textsc{TypedFields}\xspace}

\begin{document}
\setlength{\emergencystretch}{1em}

\title{Structured Feedback Improves Repair in an LLM Agent Loop}

\author{Jaideep Ray}
\email{jaray@acm.org}
\affiliation{\institution{Independent Researcher}\city{}\country{}}

\author{Ankit Goyal}
\email{ankit@goyalankit.com}
\affiliation{\institution{Independent Researcher}\city{}\country{}}

\begin{abstract}
LLM agents often retry after external validation rejects a candidate, but the
interface between validation and the next model call remains underspecified. We
introduce \vh, a code-controlled agent loop in which models generate candidates
while external validators control acceptance, budgets, and traces. We use it to
compare raw diagnostics with feedback that identifies the failure location,
observed value, and admissible alternatives.

Across 50 paired TextWorld games under a four-call cap, feedback containing all
three fields improves terminal success by 44 percentage points for
Qwen2.5-Coder-14B (95\% paired-bootstrap interval 28--60) and by 42 percentage
points for Llama-3.1-8B (95\% interval 28--56). Ablations locate most of the
gain in the admissible alternatives:
feedback containing only the location and observed value remains near the raw
diagnostic baseline. Presenting the complete repair information in prose instead
of a keyed JSON record yields nearly the same success, providing no evidence
that JSON syntax itself improves repair. The ordering persists across the tested
call budgets and one sampled-decoding setting. These results support validators
that return specific repair information rather than only rejection messages.
\end{abstract}


\maketitle

\section{Introduction}

We use three terms throughout. A \emph{model} is the LLM called to generate one
candidate action or artifact. A \emph{validator} is external code that checks a
candidate and returns either pass or a failure description. An \emph{agent loop}
is the bounded sequence model $\rightarrow$ validator $\rightarrow$ feedback,
repeated until validation passes or the call budget is exhausted. In software
development, such a loop may write a patch or plan, invoke a test or execution
environment, and react to failure. ReAct, Reflexion, Self-Refine, and
self-debugging establish the value of iterative feedback
\cite{yao2023react,shinn2023reflexion,madaan2023selfrefine,chen2024selfdebug}.
Coding agents also benefit from purpose-built interfaces
\cite{yang2024sweagent}. Yet one interface remains underspecified: \emph{what
exactly should a validator tell the next model call?}

Suppose command 1 in a plan is invalid. A raw retry sees ``Command 1 is not
admissible.'' A repair interface can additionally expose the observed command
and the commands admissible at that state. This may help because the information
supports choosing a replacement, because the failure location is explicit, or
because named fields are easier to parse. We compare these explanations through
ablations.

We use \vh, a code-as-harness prototype, to ask:
\begin{description}
  \item[RQ1.] Does structured verifier feedback with expected alternatives improve repair
  over raw diagnostics under equal compute?
  \item[RQ2.] Is any gain caused by the keyed representation, the repair values,
  or both?
\end{description}

We test both questions across two models, several call budgets, and sampled
decoding. We also use a small HumanEval experiment to check when the mechanism
does not apply.

On TextWorld, feedback with the location, observed value, and admissible
alternatives works much better than a raw diagnostic. The same repair values
work similarly in prose and a keyed record, while location without alternatives
is not enough. We provide the harness and traces needed to reproduce these
comparisons.

\section{Background and System}

Reflexion stores verbal feedback, while Self-Refine lets an LLM produce and use
its own critique \cite{shinn2023reflexion,madaan2023selfrefine}. Execution-guided
methods reuse failed programs and test messages \cite{chen2024selfdebug}.
SWE-bench and SWE-agent show the importance of executable evaluation and
agent--computer interfaces \cite{jimenez2024swebench,yang2024sweagent}. \vh
studies what a deterministic validator should return to the next LLM call.

\subsection{Code as the Control Plane}

\begin{figure}[t]
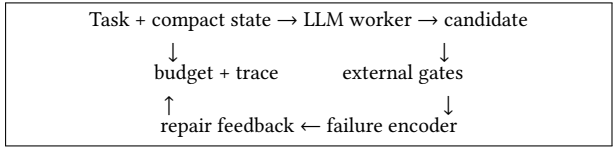

\centering
\fbox{\begin{minipage}{0.92\columnwidth}
\centering\small
Task + compact state $\rightarrow$ LLM worker $\rightarrow$ candidate\\[2pt]
$\downarrow$\hspace{98pt}$\downarrow$\\[-2pt]
budget + trace\hspace{24pt}external gates\\[2pt]
$\uparrow$\hspace{101pt}$\downarrow$\\[-2pt]
repair feedback $\leftarrow$ failure encoder
\end{minipage}}
\caption{LLMs generate leaf candidates; code controls state, validation,
acceptance, budgets, and traces.}
\Description{A loop from task and compact state to an LLM worker, candidate,
external gates, failure encoder, and repair feedback.}
\label{fig:loop}
\end{figure}

\vh separates generation from acceptance (Fig.~\ref{fig:loop}). A principal
orchestrator builds a compact state pack and dispatches bounded LLM workers.
Each worker returns a structured candidate. External gates check schema,
artifacts, and environment behavior. Only gates accept a candidate; a worker's
self-assessment cannot do so. Every prompt, raw response, parsed output, gate
result, model setting, and elapsed time is persisted before the next call.

Code is preferable here to prose as the \emph{control plane}. A prompt-only
loop asks the model to remember workflow state, count calls, interpret its own
completion, and carry an expanding history in context. Those duties compete
with candidate generation, and model-declared completion cannot independently
validate the model's output. Code instead keeps durable state outside the
prompt, enforces budgets and acceptance gates deterministically, and permits the
same tasks and failures to be replayed across policies. The model receives only
the compact task state and current repair feedback. We use this architecture to
keep the experiment controlled. We compare feedback policies; we do not compare
code orchestration with prompt-only orchestration.

The failure encoder maps gate-specific details to a common interface. For an
invalid TextWorld action it can expose a stable label, the command index, the
admissible actions in environment order, and the rejected action. Oracle-blind
evaluation excludes hidden post-hoc outcomes from repair feedback.

\subsection{Feedback Policies}

All policies receive the same task state, generation instructions, gate, model,
and call cap. The context pack names the policy, but contains no repair feedback
on the first call. After a failed candidate, the policies return different
feedback (Table~\ref{tab:policies}).
Their names describe the feedback they return:
\rawdiag means \emph{Raw Diagnosis} and adds the validator's original error
without explicit repair fields; \samenl means \emph{Same Information in Natural
Language} and adds the location, observed value, and expected alternatives in
prose; \locobs means \emph{Location and Observation} and uses a stable failure
label plus named JSON fields for the location and observed value, but no
alternatives; and \typedfields means \emph{Typed Repair Fields} and adds an
expected field containing the alternatives. Here, \emph{typed} refers to the
stable label and named fields; the saved payload serializes each field value as
a string.

\begin{table}[t]
\caption{Feedback controls used in the primary experiment.}
\label{tab:policies}
\small
\begin{tabular}{@{}p{0.22\columnwidth}p{0.70\columnwidth}@{}}
\toprule
Policy & Feedback after failure \\
\midrule
\rawdiag & Original validation error; no explicit repair fields. \\
\samenl & Location, expected alternatives, and observed value in prose, without
typed labels or keys. \\
\locobs & Failure label plus named location and observed fields; expected
alternatives omitted. \\
\typedfields & Failure label plus named location, expected, and observed fields. \\
\bottomrule
\end{tabular}
\end{table}

\samenl and \typedfields contain the same location, expected, and observed
values. \typedfields additionally has a stable failure label and JSON keys, so
their contrast tests whether that keyed presentation adds an advantage beyond
the matched repair values. \locobs and \typedfields use the same keyed format
and differ by the expected field, isolating the alternatives. \rawdiag versus
\samenl measures the value of adding all three repair values to the raw message.
Here, \emph{structured} means explicit repair components---location, expected
alternatives, and observation---whether rendered as prose or a keyed record.

\subsection{One Failure, Four Feedback Policies}

Consider a real failure from TextWorld game 20261052. The model returns
\texttt{[go east, open chest, take spork]}. After \texttt{go east}, the chest
is already open, so the second command fails. The validator records the location
\texttt{commands[1]}, the observed value \texttt{open chest}, and 12 admissible
commands. Below, $[\ldots]$ abbreviates the same alternatives in both policies.
All four retries also include the rejected plan and common output instructions.
We omit those elements and the policy wrapper, and show only the diagnostic
payload.

\begin{description}
  \item[\rawdiag] ``Raw validation message from textworld: Command 1 is not
  admissible in the current game state.''
  \item[\samenl] ``The rejected value was at
  \texttt{LeafOutput.answer.}\allowbreak\texttt{commands[1]}. The
  validator observed \texttt{open chest}. It expected
  [\texttt{close type 4 chest}, $\ldots$, \texttt{take spork from type 4
  chest}].''
  \item[\locobs] \texttt{typed\_failure=\{}\\
  \texttt{"label":"textworld.action\_invalid",}\\
  \texttt{"location":"LeafOutput.answer.}\\
  \texttt{commands[1]",}\\
  \texttt{"observed":"open chest"\}}
  \item[\typedfields] The same keyed record as \locobs, plus an
  \texttt{expected} string containing [\texttt{close type 4 chest}, $\ldots$,
  \texttt{take spork from type 4 chest}].
\end{description}

\locobs says where the plan failed but gives no replacement commands. \samenl
and \typedfields include the same location, observation, and alternatives; one
uses prose, while the other uses named fields and adds the failure label.

We use \emph{repair} to mean a new model candidate produced after a validator
rejects an earlier candidate. The new candidate receives the failure feedback
but must still pass the external validator; the model does not accept its own
answer. In this trace, \typedfields tells the model that
\texttt{commands[1]} contains the invalid action \texttt{open chest} and lists
\texttt{take spork from type 4 chest} as an admissible replacement. The next
model call returns \texttt{[go east, examine type 4 chest, take spork from type
4 chest]}. TextWorld executes this revised plan, reaches the goal, and the agent
loop stops with an accepted answer. The feedback helps by turning a general
rejection into a specific change that the model can make and the validator can
check.

\section{Study Design}

\textbf{Primary tasks.} We generate 50 fixed TextWorld games
\cite{cote2018textworld}, with seeds 20261051--20261100. Each has three rooms,
four objects, and a two-step quest. A candidate is a JSON plan of at most four
commands, executed from fresh state. Success requires a terminal win. Invalid
actions return their index and up to the first 12 admissible commands in the
environment's deterministic order; we neither rank nor randomize this list.
The primary matrix is $50$ games $\times$ four policies $\times$ two models,
or 400 rows.

\textbf{Models and decoding.} We use Qwen2.5-Coder-14B-Instruct-AWQ
\cite{hui2024qwen25coder} on an NVIDIA L4 and
Meta-Llama-3.1-8B-Instruct-AWQ-INT4
\cite{grattafiori2024llama3} on a T4, served by vLLM
\cite{kwon2023vllm}. Primary decoding is greedy with
512 output tokens and a four-call cap. Model conditions are never mixed across
serving backends.

\textbf{Robustness lanes.} First, on the same 15-game subset and both models,
we compare \rawdiag and \typedfields at budgets 2, 4, 6, and 8. Budget-4 rows
are reused from the primary matrix. Second, Qwen runs 20 games at temperature
0.3, top-$p$ 0.9, and three inference seeds; this lane compares \rawdiag,
\samenl, and \typedfields. Third, a 15-task HumanEval
\cite{chen2021codex} scope check exposes one
deterministically selected public assertion to the repair gate while retaining
the complete official tests for post-hoc success. These lanes contribute 480
additional rows, for 880 total and 2,652 realized LLM calls.

\textbf{Analysis.} Games are paired by seed. Primary success differences use
10,000 paired bootstrap samples and two-sided exact McNemar tests. We apply Holm
correction to four planned contrasts within each model. Sampled-decoding
intervals resample 20 game clusters, retaining all three repeats; $p$-values use
100,000 game-level sign flips and Holm correction. The budget and HumanEval
samples are small, so we report counts and intervals and do not treat a
$p$-value threshold as decisive. Saved-transcript whitespace counts are a prompt
size proxy, not tokenizer or billing tokens.

\textbf{Reproducibility.} The artifact contains the source, deterministic task
rules, 880 row-level results, 2,652 call traces, model settings, and the script
used to regenerate the tables and statistical tests.

\section{Results}

\subsection{Structured Feedback Improves TextWorld Success}

Table~\ref{tab:primary} reports terminal success. \typedfields improves over
\rawdiag by 44 points for Qwen (95\% interval 28--60; Holm-adjusted exact
$p=3.15\times10^{-5}$) and 42 points for Llama (28--56;
$p=3.81\times10^{-6}$). The increase appears for both the coder model and the
smaller general model.

\begin{table}[t]
\caption{Primary TextWorld results: 50 paired games per model, budget 4.}
\label{tab:primary}
\small
\begin{tabular}{@{}lrrrr@{}}
\toprule
& \multicolumn{2}{c}{Qwen-14B/L4} & \multicolumn{2}{c}{Llama-8B/T4} \\
Policy & Solved & Calls & Solved & Calls \\
\midrule
\rawdiag & 14 (28\%) & 164 & 8 (16\%) & 179 \\
\locobs & 18 (36\%) & 155 & 9 (18\%) & 174 \\
\samenl & 35 (70\%) & 147 & 29 (58\%) & 161 \\
\typedfields & \textbf{36 (72\%)} & \textbf{130} & \textbf{29 (58\%)} & \textbf{149} \\
\bottomrule
\end{tabular}
\end{table}

First-call terminal wins are 8, 11, 9, and 9 for Qwen and 6, 8, 6, and 7 for
Llama under \rawdiag, \locobs, \samenl, and \typedfields, respectively. Typed
and raw therefore differ by one first-call win for each model, compared with
final gaps of 22 and 21 wins. Most of the final difference appears after retries,
although the policy name in the initial context prevents a byte-identical
first-call comparison.

\subsection{Which Feedback Fields Matter?}

\samenl improves over \rawdiag by 42 points for both models, with no
baseline-only successes (Table~\ref{tab:causal}). \typedfields differs from the
repair-value-matched \samenl by only +2 points for Qwen and 0 for Llama; both
intervals include zero and both adjusted $p$-values are 1. We therefore find no
success advantage for the keyed record and failure label over prose containing
the same repair values.

\locobs omits admissible alternatives and performs close to \rawdiag. Adding
those alternatives increases success by 36 points for Qwen and 40 for Llama.
Final invalid-action episodes fall from 35 to 2 (raw to same-information prose)
for Qwen and from 41 to 4 for Llama. The location identifies the failed command;
the alternatives give the model valid replacements.

\begin{table}[t]
\caption{Paired contrasts for Qwen2.5-Coder-14B-Instruct-AWQ and
Meta-Llama-3.1-8B-Instruct-AWQ-INT4. CIs are paired bootstrap; $p_H$ is
Holm-adjusted exact McNemar within model.}
\label{tab:causal}
\small
\begin{tabular}{@{}llrrr@{}}
\toprule
Model & Treatment vs. baseline & $\Delta$ & 95\% CI & $p_H$ \\
\midrule
Qwen & SameNL vs. raw & +42 & [28,56] & $3.8\!\times\!10^{-6}$ \\
Qwen & Typed vs. SameNL & +2 & [$-$8,12] & 1.0 \\
Qwen & Typed vs. LocObs & +36 & [20,52] & $2.4\!\times\!10^{-4}$ \\
Llama & SameNL vs. raw & +42 & [28,56] & $3.8\!\times\!10^{-6}$ \\
Llama & Typed vs. SameNL & 0 & [$-$12,12] & 1.0 \\
Llama & Typed vs. LocObs & +40 & [26,54] & $2.2\!\times\!10^{-5}$ \\
\bottomrule
\end{tabular}
\end{table}

\typedfields also uses somewhat fewer calls at similar terminal success. It uses 17
fewer calls than \samenl for Qwen (paired mean $-0.34$ calls/game, interval
$-0.60$ to $-0.06$) and 12 fewer for Llama ($-0.24$, $-0.44$ to $-0.04$).
Average retry prompts are similar: 701 versus 716 words for Qwen and 706 versus
722 for Llama. Total realized prompt words are 14\% and 10\% lower for typed
feedback because it makes fewer calls; individual retry prompts have similar
lengths.

\subsection{Budgets and Sampled Decoding}

The difference grows between budgets 2 and 4 rather than appearing only at call
4 (Table~\ref{tab:budget}). On 15 paired games, typed-minus-raw gains rise from
20 and 27 points at budget 2 to 47 points for both models at budget 4. Qwen then
stays near 47--53 points; Llama reaches 60 points at budgets 6 and 8. \rawdiag
stays at 4/15 for Qwen and 2/15 for Llama from budgets 4 through 8. Additional
calls help \typedfields more than \rawdiag.

\begin{table}[t]
\caption{Budget sensitivity on the same 15 games. Cells are raw/typed wins.}
\label{tab:budget}
\small
\begin{tabular}{@{}lrrrr@{}}
\toprule
Model & $B=2$ & $B=4$ & $B=6$ & $B=8$ \\
\midrule
Qwen & 5/8 & 4/11 & 4/11 & 4/12 \\
Llama & 1/5 & 2/9 & 2/11 & 2/11 \\
\bottomrule
\end{tabular}
\end{table}

Sampled Qwen decoding gives the same ordering. Across seeds 3101--3103,
\rawdiag solves 6, 6, and 5 of 20 games; \samenl solves 14, 13, and 14; and
\typedfields solves 14, 16, and 14. Clustered across games, typed versus raw is
+45 points (95\% interval 23--67; Holm $p=.0053$), while typed versus
repair-value-matched prose is +5 points ($-3$ to 15; $p=.499$). The primary
ordering therefore also appears with sampled decoding.

\subsection{When the Validator Cannot Expose the Hidden Failure}

The HumanEval experiment tests when repair feedback can help, rather than
serving as another positive benchmark. For each task, the validator runs one
visible test. If the first answer fails that test, the harness can return the
failure to the model and request a fix; we call this a \emph{repair
opportunity}. The complete HumanEval test suite remains hidden and is used only
to score the final answer.

For Qwen, the first answer passes the visible test on all 15 tasks, so the
harness never receives a repair opportunity and every policy stops after one
call. Fourteen answers pass the hidden suite. The remaining answer passes the
visible test but fails a hidden test, so no feedback policy can detect or repair
it. All policies therefore finish at 14/15.

Structured feedback can help only when the validator exposes a useful failure.
It cannot fix behavior that the validator does not test, and passing one visible
test does not guarantee hidden correctness. We therefore treat HumanEval as a
scope check, not as evidence that the method improves code repair in general.

\section{Implications}

\textbf{Validator output.} A validator should report where the failure occurred,
what it observed, and what it expected when that information is available. In
software, these fields may be a file and line, failing assertion, expected type
or behavior, and observed exception. When a test cannot enumerate alternatives,
the validator should report only what it knows.

\textbf{Representation.} Named JSON fields can simplify logging, routing, and
cross-gate tooling, and our keyed condition uses somewhat fewer calls. Prose
with the same repair values, however, reaches nearly the same success rate in
this study.

\textbf{Call budgets.} In the 15-game budget subset, \rawdiag is flat from four
through eight calls. Structured feedback continues improving Llama through six
calls. Extra calls are more useful when the retry receives new information.

\section{Limitations}

This is a focused study. The main experiment uses 50 generated TextWorld
games and two quantized models served with vLLM. Sampled decoding covers only
Qwen and three inference seeds. The HumanEval check has 15 tasks. We have not
tested repository-scale repair or production agent systems. HumanEval also
shows a basic limit: feedback cannot help when the visible validator does not
detect the hidden failure.

Some design choices may affect the results. We cap admissible actions at 12 and
keep the order returned by TextWorld. We did not test longer, ranked, or
randomized lists. The prose and keyed policies contain the same location,
expected, and observed values, but the keyed policy also includes a stable
failure label and different framing text. Their comparison is therefore not a
pure punctuation-only contrast. The first-call context also includes the policy
name; typed and raw differ by one initial win in each model, but this still
prevents a strictly identical pre-repair prompt. Prompt size is measured with
transcript words rather than model tokens. Real validators can be incomplete or flaky. One
serving error in this study was rerun with the same settings, and only the
replacement run is included.

\section{Conclusion}

We asked what a validator should return to the next model call after a failed
candidate. For RQ1, feedback with a failure location, observed value, and
admissible alternatives improves terminal TextWorld success under the same
four-call cap: from 14/50 to 36/50 for Qwen and from 8/50 to 29/50 for Llama. On
the 15-game budget subset, increasing \rawdiag from four to eight calls does not
solve additional games. The HumanEval check shows an important condition: a
repair policy cannot act on a failure that its visible validator does not expose.

For RQ2, the large gain comes from the repair values rather than an observed
advantage for the keyed representation. \samenl improves over \rawdiag by 42
points for both models. Its success differs from \typedfields by two points for
Qwen and zero for Llama. In contrast, \locobs omits admissible alternatives and
stays near \rawdiag. Within this experiment, the alternatives account for most
of the improvement.

For agent design, a validator should return what it knows about the failed
candidate, especially valid alternatives when they can be enumerated. Named JSON
fields may still be useful for logging and routing, but these results do not show
that the keyed format itself improves model reasoning.

\balance
\bibliographystyle{ACM-Reference-Format}
\bibliography{references}

\end{document}